%% file: main.tex
\documentclass[twocolumn,secnumarabic,amssymb, nobibnotes, aps, prd, preprintnumbers, showkeys]{revtex4-2}
\usepackage{graphicx, subcaption, booktabs, braket, mathtools, tikz, textcomp}
\newcommand{\var}{\mathrm{var}}
\newcommand{\oo}{\mathrm{O}}

\graphicspath{{plots/}}
\begin{document}

\preprint{DESY-24-089}

\begin{abstract}
This study explores the application of a two-level algorithm to enhance the signal-to-noise ratio of glueball calculations in four-dimensional $\mathrm{SU(3)}$ pure gauge theory. Our findings demonstrate that the statistical errors exhibit an exponential reduction, enabling reliable extraction of effective masses at distances where current standard methods would demand exponentially more samples. 
However, at shorter distances, standard methods prove more efficient due to a saturation of the variance reduction using the multi-level method. We discuss the physical distance at which the multi-level sampling is expected to outperform the standard algorithm, supported by numerical evidence across different lattice spacings and glueball channels. 
Additionally, we construct a variational basis comprising 35 Wilson loops up to length 12 and 5 smearing sizes each, presenting results for the first state in the spectrum for the scalar, pseudoscalar, and tensor channels.
\end{abstract}

\keywords{Lattice QCD, Algorithm, Glueballs}

\title{Exponential Error Reduction for Glueball Calculations Using a Two-Level Algorithm in Pure Gauge Theory}

\author{Lorenzo Barca}
\email{lorenzo.barca@desy.de}
\affiliation{John von Neumann-Institut f\"ur Computing NIC, Deutsches Elektronen-Synchrotron DESY, Platanenallee 6, 15738 Zeuthen, Germany}

\author{Francesco Knechtli}
\email{knechtli@uni-wuppertal.de}
\affiliation{Department of Physics, University of Wuppertal, Gaußstrasse 20, 42119 Germany}

\author{Sofie Martins}
\email{martinss@imada.sdu.dk}
\affiliation{$\hbar$QTC \& IMADA, University of Southern Denmark, Campusvej 55, 5230 Odense M, Denmark}

\author{Michael~Peardon}
\email{mjp@maths.tcd.ie}
\affiliation{School of Mathematics, Trinity College Dublin, Ireland}

\author{Stefan Schaefer}
\email{stefan.schaefer@desy.de}
\affiliation{John von Neumann-Institut f\"ur Computing NIC, Deutsches Elektronen-Synchrotron DESY, Platanenallee 6, 15738 Zeuthen, Germany}

\author{Juan Andr\'es Urrea-Ni\~{n}o}
\email{urreanino@uni-wuppertal.de}
\affiliation{Department of Physics, University of Wuppertal, Gaußstrasse 20, 42119 Germany}

\maketitle
\input{sections/sec1}
\input{sections/sec2}

\input{sections/sec3}

\input{sections/sec4}

\appendix
\input{appendices/appendix1}

\bibliography{bibliography/bibliography}
\bibliographystyle{unsrt}

\end{document}

%% file: sections/sec1.tex
\section{Introduction}\label{Section:sec1}

One consequence of confinement of quarks and gluons by the strong force in
quantum chromodynamics (QCD) is the possible existence of glueballs
\cite{Fritzsch:1972jv, Fritzsch:1975tx}, states predominantly made of 
gluons.  These hypothetical composites have been studied widely within a range 
of theoretical frameworks, such as the bag model \cite{PhysRevD.9.3471,
Jaffe:1975fd, Jaffe:1985qp, Chanowitz:1982qj}, flux-tube model
\cite{PhysRevD.31.2910, Iwasaki:2003cr} and QCD sum rules \cite{Coyne:1980zd,
Narison:1984hu, Huang:1998wj, Li:2021gsx} and have been investigated from first
principles since the development of lattice QCD methods \cite{Berg:1982kp,
Bali:1993fb, Morningstar:1999rf, Chen:2005mg, Athenodorou:2020ani}. Finding
experimental evidence for glueballs remains a challenge \cite{PANDA:2009yku} 
as the lightest
candidates suggested by theoretical studies are rather massive, approaching
twice the mass of a nucleon and have quantum numbers in common with some
quark-model mesons.  As a result, physical states at this energy scale are
mixtures of conventional isoscalar mesons and glueballs and have large
phase-space for decay to multiple light mesons 
\cite{Petrov:2022ipv, Klempt:2021wpg, Amsler:1995td, Donoghue:1980hw,
Kiesewetter:2010ze}. Very little is known about the decay processes of the
glueball in QCD, so reliable interpretation of resonances in collider
experiments as glueballs is challenging.  There have been a number of
experimental observations of candidates \cite{Crede:2008vw}, mostly identified 
as an over-population of resonances compared to quark model predictions, but 
their link to QCD remains unresolved. 
Recent experimental confirmation of pseudoscalar quantum numbers
for the X(2370) \cite{BESIII:2023wfi} have given fresh incentive to these 
studies.

Lattice QCD calculations can provide a great deal of information to guide these
searches from first principles. Ideally, the spectrum is computed on
the lattice, including the dynamics of the quark and gluon fields 
\cite{Gregory:2012hu, Hart:2001fp, Bali:2000vr, Hart:2006ps, Gui:2012gx, 
  PhysRevLett.111.091601, PhysRevD.100.054511} which causes
both mixing of quark-model mesons with the glueballs and their strong decays.  
Then, widths of the resulting resonances can be studied following ideas 
introduced by L\"uscher \cite{Luscher:1986pf} which relate the energy spectrum 
in a finite volume to certain scattering phase shifts. Developing this framework beyond two-body
elastic scattering is a very active research topic \cite{Briceno:2012rv}, but 
the 
input which lattice calculations must provide remains the accurate 
determination of a complete low-energy spectrum of QCD in a finite volume. 
This can only be computed in practice by determining two-point correlation 
functions between a broad range of creation and annihilation operators 
resembling glueballs, quark-model isoscalar mesons, two-pion states, and so on.  

One difficulty with these calculations arises from use of importance-sampling
Monte Carlo estimates. The spectrum is extracted from the Euclidean-time
dependence of correlation functions, which fall exponentially with a rate
governed by the energies of the states under investigation. In glueball
calculations, the relevant scale is typically beyond 1.5 GeV. The Monte Carlo
estimators have high statistical variance, roughly independent of the Euclidean
time separation. As a result, the signal-to-noise ratio falls rapidly 
\cite{Parisi:1983ae, Lepage:1989hd} and very large statistical ensembles, 
typically with $\mathcal{O}(10^5)$ samples, are needed for high precision. 

This study tests and models the behavior of improved Monte Carlo methods
offering better signal-to-noise performance. Our investigation is restricted to
the pure Yang-Mills $\mathrm{SU(3)}$ theory; there is active research 
\cite{Ce:2016idq, Ce:2016ajy, Giusti:2017ksp, Ce:2017ndt, Ce:2019yds} into 
extensions of these techniques to QCD, including the dynamics of the quark 
fields outside the scope of this paper. Calculations with fermions present 
significant technical challenges as the resulting path integrals cannot be 
computed directly by Monte Carlo.  Locality of the Yang-Mills action along 
with bosonic statistics of gluons allows
factorization of the observables of the lattice field theory into sub-volume
integrals, which can be sampled independently by drawing random field variables
from the appropriate conditional distributions and constructing full
observables with smaller variance from products of these independent samples.
The first use of this factorization was the 
\textit{multi-hit algorithm} \cite{PARISI1983418}, introduced in computations 
of the string tension. The algorithm was
subsequently extended to \textit{multi-level} methods
\cite{Luscher:2001up}, achieving exponential
error reduction for certain observables.  Later work 
\cite{Meyer:2002cd, Meyer:2003hy, Hasenbusch:2004yq, Majumdar:2014cqa}
investigated these schemes
for glueball calculations where the temporal extent of the lattice is divided
into nested subdomains and expectation values estimated by hierarchical Markov
chain Monte Carlo calculations inside these layers. 

The paper extends earlier work \citep{Barca:2023arw} testing these schemes in 
detail and is organized as 
follows. Section 2 reviews the multi-level sampling algorithm, focussing on 
determinations of glueball correlation functions. Section 3 presents the 
findings of our calculation, including detailed analysis of the performance 
of the method when using a large basis of operators suitable for spectrum 
calculations based on solving a generalized eigenvalue problem (GEVP)
\cite{MICHAEL1983433, Blossier:2009kd, Bulava:2011yz}. These 
techniques are the key
to reliable studies when mixing with quarkonia or decays into multiple 
hadrons are investigated. 
Comparisons with state-of-the-art calculations are made, and statistical
precision is studied in detail. Our conclusions are presented in Section 4. 

%% file: sections/sec2.tex
\section{The multi-level sampling algorithm}\label{Section:sec2}
Energy eigenstates can be investigated using two-point correlation functions
of quantum fields on an Euclidean lattice. Their efficient determination in 
Monte Carlo calculation using a multi-level algorithm proceeds as follows.
\subsection{Algorithmic details}\label{Section:sec2_subsec1}
The correlation function of two operators $\mathrm{O}(t_1)$ and $\bar{\mathrm{O}}(t_0)$ is defined as
\begin{align}\label{correlator_continuum}
\begin{split}
C(t_1, t_0) &\equiv  \langle \oo(t_1) \bar{\oo}(t_0) \rangle  \\
&=
\frac{1}{\mathcal{Z}} \int dU ~e^{-S(U)} \oo(U, t_1) \bar{\oo}(U, t_0)~,
\end{split}
\end{align}
where $\mathcal{Z}$ is the partition function, and $S$ is the Euclidean action.
This integral can be estimated numerically with Monte Carlo techniques, 
by generating an ensemble of $N$ gauge configurations 
$\{U_1, ~U_2, ~\dots U_{N}\}$, drawn from a sample distribution with 
probability $P(U_i) \propto e^{-S(U_i)}$. We then estimate the correlation by 
computing
\begin{equation}\label{correlator_mc}
	\widehat{C}(t_1,t_0)_{\mathrm{standard}} = \frac{1}{N} \sum_{i=1}^N \oo(U_i, t_1) \bar{\oo}(U_i, t_0)~
\end{equation}
with the associated standard error
\begin{equation}\label{error_standard}
	\sigma_C\left(t_1, t_0; N \right)
	=
	\sqrt{\frac{\var\left(\oo(t_1)\bar{\oo}(t_0)\right)}{N}}~,
\end{equation}
which depends on the decorrelated sample size $N$ and variance
\begin{equation}\label{variance_standard}
	\var \left(\oo(t_1)\bar{\oo}(t_0) \right)= 
	\langle \left(\oo(t_1) \bar{\oo}(t_0) \right)^2\rangle
	-
	\langle \oo(t_1) \bar{\oo}(t_0) \rangle^2~.
\end{equation}
Multi-level simulations, \cite{PARISI1983418, Luscher:2001up, Meyer:2002cd} 
were introduced after observing the path integral of the correlation function 
in a pure gauge theory factorizes into subdomains 
\begin{align}
\begin{split}
C(t_1, t_0)=&
\int dU_B ~\bigg\{\dfrac{e^{-S_B(U_B)}\mathcal{Z}_1(U_B)\mathcal{Z}_2(U_B)}{\mathcal{Z}} \times
\\&\qquad\int dU^{(1)} ~\dfrac{e^{-S_1(U^{(1)}|U_B)}}{\mathcal{Z}_1(U_B)} \oo(U^{(1)}, t_1) \\&\qquad\int dU^{(2)} ~\dfrac{e^{-S_2(U^{(2)}|U_B)}}{\mathcal{Z}_2(U_B)} \bar{\oo}(U^{(2)}, t_0)\bigg\}\\
=&~\langle [\oo(U^{(1)},t_1)] ~ [\bar{\oo}(U^{(2)},t_0)]\rangle~,
\label{PI-fact}
\end{split}
\end{align}
where we are using, in addition to the established expectation $\langle \ldots \rangle$, the sub-average as defined in \cite{Luscher:2001up}
\begin{equation}
[O(U^{(r)},t)] = \int dU^{(r)} ~\dfrac{e^{-S_r(U^{(r)}|U_B)}}{\mathcal{Z}_r(U_B)} \oo(U^{(r)}, t)~,
\end{equation}
with
\begin{equation}
\mathcal{Z}_{r}(U_B) = \int dU^{(r)} e^{-S_{r}(U^{(r)}| U_B)}\,.
\end{equation}
Here, we denote $U_B$ as gauge fields on fixed boundaries, while $U^{(1)}$ and $U^{(2)}$ are gauge fields whose dynamics are determined by the conditional actions $S_1$ and $S_2$. The superscripts of $U^{(1)}$ and $U^{(2)}$ refer to the two different regions. 

This decomposition splits the lattice into two independent domains depending on the fixed boundary $B$ separating them. The path integrals on the subdomains can then be approximated by simulating only the subdomain with fixed boundary conditions. 
In practice, we initially generate a regular Monte Carlo chain of $N_0$ configurations, denoted as $U_i$ for $i= 1,\ldots, N_0$.
These are referred to as \textit{level-0} configurations. 
The fields restricted to $B$ are subsequently distributed according to the marginal probability density in the $ U_B$ integration in Eq.~\eqref{PI-fact}.

\begin{figure}[tp!]
\includegraphics[width=0.5\textwidth]{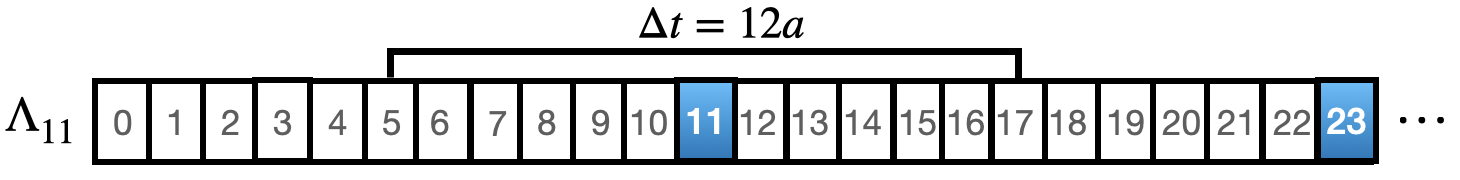}
	\caption{Example of the sub-lattice decomposition $\Lambda_{11}$ (ellipses represent $t/a\ge 24$), 
	where the dynamical regions are eleven lattice spacings wide. The angled connecting line visualizes the correlation between operators measured in two different dynamical regions. For instance, in this case, one at $t_0=5a$ and the other operator at $t_1=t_0+\Delta t=17a$.}
\label{Figure:dt1plus11}
\end{figure}

From each of these level-0 configurations, we generate a sub-chain of configurations updated in each sweep only on partial domains but left unchanged on the remaining temporal domains. These sub-chains are generated with fixed boundary conditions, drawn from the conditional probability density in the $U^{(r)}$-integrals in Eq.~\eqref{PI-fact}.
We refer to the updated domains as the \textit{dynamical regions}, while the other time slices are the \textit{fixed regions}, which separate two different dynamical ones. 
We use the notation $\Lambda_k$ to refer to the specific sub-lattice decomposition in dynamical and frozen regions of the sub-chains. In Fig.~\ref{Figure:dt1plus11}, there is one example of sub-lattice decompositions adopted in this work.
The blue-shaded regions at $t=11a$, $23a$, $\dots$ ($35a$, $47a$) are the fixed regions, consisting of only one time slice each, and the other time slices are the dynamical regions, where the fields $U_i$ are updated along the level-1 chain.
The temporal gauge links connecting the fixed with the dynamical regions are updated. We label these \textit{level-1} configurations, $U_{ij}$ using a second index $j\in 1, \ldots, N_1$ to indicate the trajectory time along the sub-chains, while $i$ indicates the initial level-0 configuration.
Thus we generate $N_0 \times N_1$ gauge configurations, labelled $U_{ij}$. 
Using the Wilson plaquette action, fixed regions comprised of a single time slice are sufficient to separate the dynamical regions such that gauge updates are independent, whereas for improved gauge actions which include larger Wilson loops such as the L{\"u}scher-Weisz, the width of the frozen regions must be increased.

When the operators are in two different dynamical regions, the multi-level estimator reads
\begin{align}\label{correlator_mlvlmc}
\widehat{C}(t_1, t_0)= \frac{1}{N_0} \sum_{i=1}^{N_0}
\bigg(\frac{1}{N_1}\sum_{j=1}^{N_1}\oo_{ij}(t_1) \bigg)
\bigg(\frac{1}{N_1}\sum_{k=1}^{N_1} \bar{\oo}_{ik}(t_0) \bigg)
\\
\nonumber
\text{for } t_0, t_1 \text{ in different domains,} \hspace{3em}
\end{align}
which corresponds to a combination of two independent level-1 averages followed by a level-0 average.
We will confirm that this reduces noise sufficiently to make it equivalent to a standard measurement with $N_0 \times N_1^2$ samples up to boundary corrections. Then the error scales ideally like $\sigma_C(t_1, t_0; N_0; N_1) \sim 1/ \sqrt{N_0  N_1^2}$, as one can guess from the scaling of this estimator with $N_0$ and $N_1$. Corrections to this ideal scaling are discussed in the following subsection. 
The factorization in Eq.~\eqref{PI-fact} applies only when the operators are in two different regions.
 
When the operators are in the same region, and therefore when $t_0$ and $t_1$ belong to the same domain, the level-1 and level-0 measurements give a unique sample, and we have to employ the standard estimator
\begin{align}\label{estimate_correlator_mlvlmc_sameregion}
\widehat{C}(t_1, t_0)=
\frac{1}{N_0 N_1} \sum_{i=1}^{N_0} \sum_{j=1}^{N_1}
\oo_{ij}(t_1) \bar{\oo}_{ij}(t_0)
\\
\nonumber
\text{for } t_0, t_1 \text{ in same domain,} \hspace{1em}
\end{align}
where at best the error scales like $\sigma_C(t_1, t_0; N_0; N_1) \sim 1/ \sqrt{N_0  N_1}$ as in Eq.~\eqref{error_standard}.
\subsection{The noise-to-signal ratio}\label{nts-ratio-subsec}
The spectral decomposition of the correlation functions, estimated with Monte Carlo algorithms using Eq.~\eqref{correlator_mc},
exhibits an exponential decay of the signal over time as
\begin{align}\label{eq:exp-decrease-correlator}
\begin{split}
C(t_1, t_0) &= \langle \oo(t_1)\bar{\oo}(t_0)\rangle \\&= \sum_{n} |\bra{0}\oo(0)\ket{n}|^2 e^{-E_n(t_1-t_0)}
\\&~\stackrel{t_1\gg t_0}{\sim} ~
e^{-m(t_1-t_0)}~,
\end{split}
\end{align}
with $E_n$, the energy of a state with the quantum numbers of the zero-momentum operator $\oo$, and $m$, the ground-state mass. 
Since in our investigation $\oo(t)$ is a glueball operator, $m$ is the mass of 
the ground-state glueball. 

The variance of a glueball correlation function computed with traditional algorithms is approximately constant \cite{Lepage:1989hd, Parisi:1983ae} as the source-sink separation $t_1-t_0$ varies. Thus, its noise-to-signal ratio increases exponentially as
\begin{equation}
\dfrac{\sigma_C(t_1, t_0; N)}{C(t_1, t_0)} \sim \dfrac{1}{\sqrt{N}}e^{m(t_1-t_0)} ~,
\end{equation}
and to keep the relative precision constant at all time separations, one must record $N \sim e^{2m(t_1-t_0)}$ measurements. 
However, if the correlation function is estimated with multi-level algorithms using Eq.~\eqref{correlator_mlvlmc}, the noise-to-signal ratio becomes
\begin{equation}\label{ntos_mlvl_naive}
 \dfrac{\sigma_C(t_1, t_0;N_0; N_1)}{C(t_1, t_0)} \sim \dfrac{1}{\sqrt{N_0 N_1^2}} e^{m(t_1-t_0)},
\end{equation}
where clearly, performing the additional $N_1$ sub-measurements keeps the 
relative error fixed much more efficiently. 
Indeed, performing $N_1 \approx e^{m(t_1-t_0)}$ more sub-measurements for each operator $\mathrm{O}$ keeps the error fixed, 
as opposed to $N \approx e^{2m(t_1-t_0)}$ with the traditional algorithm. There is a clear exponential improvement.

However, the expression in Eq.~\eqref{ntos_mlvl_naive} is correct up to fluctuations arising from the boundaries, i.e., the fixed regions, but $N_1 \approx e^{m(t_1-t_0)}$ will still be enough to keep the relative error constant, as we discuss below.
In general, by taking into account fluctuations from the fixed boundaries, one can derive the expression for the variance of the multi-level correlation function \cite{GarciaVera:2016dau}, which reads
\begin{align}\label{mlvl_error}
\begin{split}
\sigma^2_C(t_1, t_0; N_0; N_1) &\approx 
\dfrac{c_0^2}{N_0 N_1^2} \\
&+ \dfrac{c_1^2}{N_0 N_1}  \left(e^{-\widetilde{m}\Delta_1} +  e^{-\widetilde{m}\Delta_0}\right)\\
&+ \frac{c_2^2}{N_0} e^{-\widetilde{m}\left(\Delta_1 + \Delta_0\right)}~,
\end{split}
\end{align}
with $c_0, c_1, c_2$ of similar order of magnitude.
In this formula, $\Delta_1$ and $\Delta_0$ denote the distance of the operators at $t_1$ or $t_0$ respectively to the nearest boundaries at $t_B$. 

Eq.~\eqref{mlvl_error} contains important information on the dependence of multi-level algorithm performance on the parameter choice. 
The time-independent term of the multi-level error $\sigma_{C}(t_1,t_0;N_0, N_1)$ scales with $1/\sqrt{N_0N_1^2}$, up to exponential fluctuations induced 
by the fixed boundaries.
The multi-level error decreases with the number of sub-measurements $N_1$ until the last term in Eq.~\eqref{mlvl_error},
which is independent of $N_1$, becomes relevant. 
Note the formula in Eq.~\eqref{mlvl_error} accounts only for fluctuations from one nearest boundary for each operator, 
while in general, one has to consider all the boundary fluctuations.

We must emphasize an important caveat: Numerically, we find that the mass in the exponent of the boundary fluctuations corresponds to twice the mass of a glueball state, $\widetilde{m} = 2m^\Gamma$, in that particular glueball channel $\Gamma=A_1^{++}, E^{++}, T_2^{++}, A_1^{-+}$.
In all glueball channels, we expect states with vacuum quantum numbers to contribute at long distances and to be suppressed at short distances due to small overlap.
Unfortunately, we do not have sufficient data to show this.
\subsection{Saturation of the variance reduction}
The multi-level error reduction is explicitly dependent on the level-1 sub-measurements and the distance of the operators from the nearest boundaries. 
In particular, there is a critical number of sub-measurements $\widetilde{N}_1$, above which the multi-level error does not reduce further with $N_1$.
This saturation point is reached when the first two terms in Eq.~\eqref{mlvl_error} become the same order of magnitude as the last.

As a result, for operators symmetrically distant from the boundaries, i.e. $\Delta_1 = \Delta_0 =  \Delta t/2$, with $\Delta t=t_1-t_0$ and $N_1 \gg \widetilde{N}_1 \sim \mathrm{exp}(m^\Gamma\Delta t)$, the multi-level error decreases exponentially like $\sim \mathrm{exp}(-m^\Gamma\Delta t)$ and consequently the relative error stays constant with increasing $\Delta t$. As noted in section \ref{nts-ratio-subsec}, we see here again the origin of the noise reduction: For the standard Monte Carlo method, we need to rescale the number of measurements with $\sim \mathrm{exp}(2m^\Gamma \Delta t)$ while now we need only $\sim \mathrm{exp}(m^\Gamma\Delta t)$ more measurements in a multi-level simulation to achieve a constant relative error. 

However, at short distance, this represents a limitation compared to the traditional error scaling. While the multi-level error is independent of $N_1$ for $N_1 \gg \tilde{N}_1$, the traditional scaling decreases as $1/\sqrt{N_1}$, modulo exponential fluctuations from the boundaries. 
Therefore, at short distance, it is better to measure the correlations between operators located in the same dynamical regions, whose error scales like the traditional scaling.
There is an intermediate distance $\Delta \tilde{t}$ where the saturated multi-level starts to outperform the standard scaling. This occurs when
 \begin{equation}\label{transition_point}
	\mathrm{exp}\left(-m^\Gamma \Delta \tilde{t} \right) \lesssim \frac{1}{\sqrt{N_1}}
 \end{equation}
is satisfied. 
Our simulations confirm this behavior, as shown in the next section.
For instance, on the ensemble \textit{gb62}, it is clear from Figs.~\ref{fig:err_fixed_t}-\ref{Figure:universal_error_scaling_continuum}
the transition point for the $E^{++}$ channel with $am^{E^{++}}\approx 0.77$ and $N_1=1000$ is between $4a \le \Delta \tilde{t} \le 6a$.

The variance of the correlation function $C(t_1, t_0)$ is clearly not the same for each $t_0$,
even at fixed $\Delta t$.
To take advantage of all the measurements 
for a given distance $\Delta t = t_1-t_0$, we consider the weighted average
\begin{equation}\label{weighted_avrg}
	\overline{C}(\Delta t) = 
	\frac{\sum_{t_0} w(t_0+\Delta t, t_0) C(t_0 + \Delta t, t_0)}
	{\sum_{t_0 } w(t_0 + \Delta t, t_0)}~,
\end{equation}
where we choose the weights $w(t_1, t_0) = 1/ \sigma^2_C(t_1, t_0;N_0; N_1)$. This weighted average selects the best correlation from either the standard correlation of operators in the same domain for short distances or multi-level correlations of operators separated equally from a fixed boundary in different domains for long distances.
There is more discussion on optimum weighting for multi-level algorithms in \cite{Kitching-Morley:2022kry}. 

Due to the weighted average, there is a transition between different error scalings:
At sufficiently long distance $\Delta t = t_1-t_0 \gtrsim \Delta \tilde{t}$, which depends on the glueball channel, c.f. Eq.~\eqref{transition_point}, 
from Eq.~\eqref{mlvl_error} we expect the variance-to-signal ratio of the weighted average in Eq.~\eqref{weighted_avrg} for a specific glueball channel $\Gamma$ to be
\begin{align}\label{ns_longdistance}
\begin{split}
\frac{\sigma^2_{\overline{C}}(\Delta t)}{\overline{C}^2(\Delta t)}
&\approx 
\frac{\tilde{c}_0^2}{N_0 N_1^2} e^{2m^\Gamma \Delta t}
+
\frac{2\tilde{c}_1^2}{N_0 N_1} e^{2m^\Gamma \Delta t/2}
+
\frac{\tilde{c}_2^2}{N_0}
\\&\hspace{11em}
\mathrm{for}
~\Delta t \gtrsim \Delta \tilde{t}~,
\hspace{1em}
\end{split}
\end{align}
where the term increasing exponentially with $\Delta t$ is divided by $N_1^2$ rather than $N_1$, thus rendering a substantial improvement at large distance.
At short distance, the noise-to-signal ratio of the weighted average is expected to scale like
\begin{equation}\label{ns_shortdistance}
\frac{\sigma^2_{\overline{C}}(\Delta t)}{\overline{C}^2(\Delta t)}
\approx 
\frac{\widetilde{a}^2_0}{N_0 N_1} e^{2m^\Gamma \Delta t}
\hspace{3.7em}
\mathrm{for}
~\Delta t \lesssim \Delta \tilde{t}~,
\hspace{1em}
\end{equation}
where $\widetilde{a}^2_0 \propto \var \left(\oo(\Delta t)\bar{\oo}(0a)\right)$, c.f. Eq.~\eqref{error_standard}.
Thus, the weighted average exhibits two different scalings at short and long distances, and the transition between the two behaviors occurs when Eq.~\eqref{transition_point} is fulfilled.

%% file: sections/sec3.tex
\section{Simulations}\label{Section:sec3}
The performance of the multi-level method was examined in several numerical 
computations of a lattice discretisation of $4D$ SU(3) Yang-Mills theory as
described in this section. 
\subsection{Details}\label{Section:sec2_subsec2}
We simulate the Wilson pure gauge action 
\begin{equation}
	S_{g} = \dfrac{\beta}{3}\sum_{n\in\Lambda}\sum_{\mu < \nu} \mathrm{Re}\{\mathrm{Tr}[1 - U_{\mu}(n)U_{\nu}(n+\hat{\mu})U^{\dagger}_{\mu}(n+\hat{\nu})U_{\nu}^{\dagger}(n)]\}
\end{equation}
with periodic boundary conditions, and we investigate $\beta=5.8$, $6.08$, $6.2$ with a physical volume kept roughly constant to $V=T\times L^3 = (3.26~\mathrm{fm}) \times (1.63~\mathrm{fm})^3$ 
using a modified version of OpenQCD \cite{openqcd, Luscher:2012av}. See Tab.~\ref{tbl:ensembles} for an overview of the ensembles analyzed in this work.
\begin{table}
	\centering
	\caption{Overview of ensembles. The tag \textit{Level-1} \textit{decomposition} denotes the size of the fixed and dynamic regions in the format: \textit{width of fixed region}+\textit{width of dynamic region}. The lattice sizes and dynamic regions are scaled to have the same physical size. Simulations are performed using the HMC with $\tau=3.0$, where the level-0 configurations are separated by 300 MDUs (molecular dynamics units). All ensembles consist of $N_0=101$ level-0 and $N_1=1000$ level-1 configurations. The scale setting is made through the Sommer scale $r_0=0.5~\mathrm{fm}$ according to \cite{Necco:2001xg}.}
	\begin{tabular}{lclclclclclclcl}
		\toprule
		Name &$~T/a~$  &   $L/a$ & ~~Level-1 decomposition~ &   ~$\beta$ & $a$ \\
		\midrule
		gb62         & 48 & ~24 & $\Lambda_{11}=1+11$		    & 6.2      & ~~$0.068~\mathrm{fm}$ \\
		gb608       & 40 & ~20 & $\Lambda_{9}=1+9$                & 6.08   & ~~$0.081~\mathrm{fm}$ \\
		gb58         & 24 & ~12 & $\Lambda_{5}=1+5$	    & 5.8	   & ~~$0.136~\mathrm{fm}$ \\
		\bottomrule
	\end{tabular}
	\label{tbl:ensembles}
\end{table}
For each $\beta$, we generate $N_0=101$ gauge configurations with the traditional HMC algorithm, i.e., the gauge configurations are updated on the entire volume. We choose well spaced configurations such that the autocorrelations are negligible. 
For each of these level-0 configurations, a new, independent Monte Carlo chain is generated using HMC,
as discussed in section~\ref{Section:sec2_subsec1}.
In particular, on the ensemble \textit{gb62} at $\beta=6.2$, with $L=24a$, $T=48a$, 
the level-1 gauge configurations are kept fixed on time slices $t = 11a,~23a,~35a,~47a$, while updated everywhere else. 
This way, there are four regions where the gauge fields are updated independently. 
This decomposition is labelled $\Lambda_{11}$ since the dynamic regions comprise $11$ lattice sites, 
corresponding to $\sim 0.75~\mathrm{fm}$ each, as seen in Fig.~\ref{Figure:dt1plus11}.
Since we are interested in the glueball spectrum, the observable $\oo(t)$ is a glueball operator,
which in pure gauge theory can be constructed from Wilson loops. 
Appendix~\ref{appendix:app1} gives more detail on constructing glueball 
operators projected onto a specific lattice irreducible representation $R^{PC}$ of the group of rotations of the cube with parity $P$ and charge conjugation $C$.
This work investigates $\Gamma \equiv R^{PC} = A_1^{++}, ~E^{++}, ~T_2^{++}$, and $A_1^{-+}$.
For each operator $\oo^{\Gamma}$ we estimate the correlation function $C^{\Gamma}(t_1, t_0)$ 
using Eq.~\eqref{correlator_mlvlmc} when the operators are in two different regions, 
while when the operators are in the same region, we estimate the correlation function 
using Eq.~\eqref{estimate_correlator_mlvlmc_sameregion}. When the two operators are in 
different dynamic regions, the error should scale according to Eq.~\eqref{mlvl_error}. 
However, if the operators are in the same region, the error is almost constant over time 
and should scale like $1/\sqrt{N_1}$, see Eq.~\eqref{error_standard}.
\subsection{Analysis of the statistical precision}
\begin{figure*}[htbp]
	\centering
	\begin{subfigure}[b]{0.85\textwidth}
		\centering
		\includegraphics[width=\textwidth]{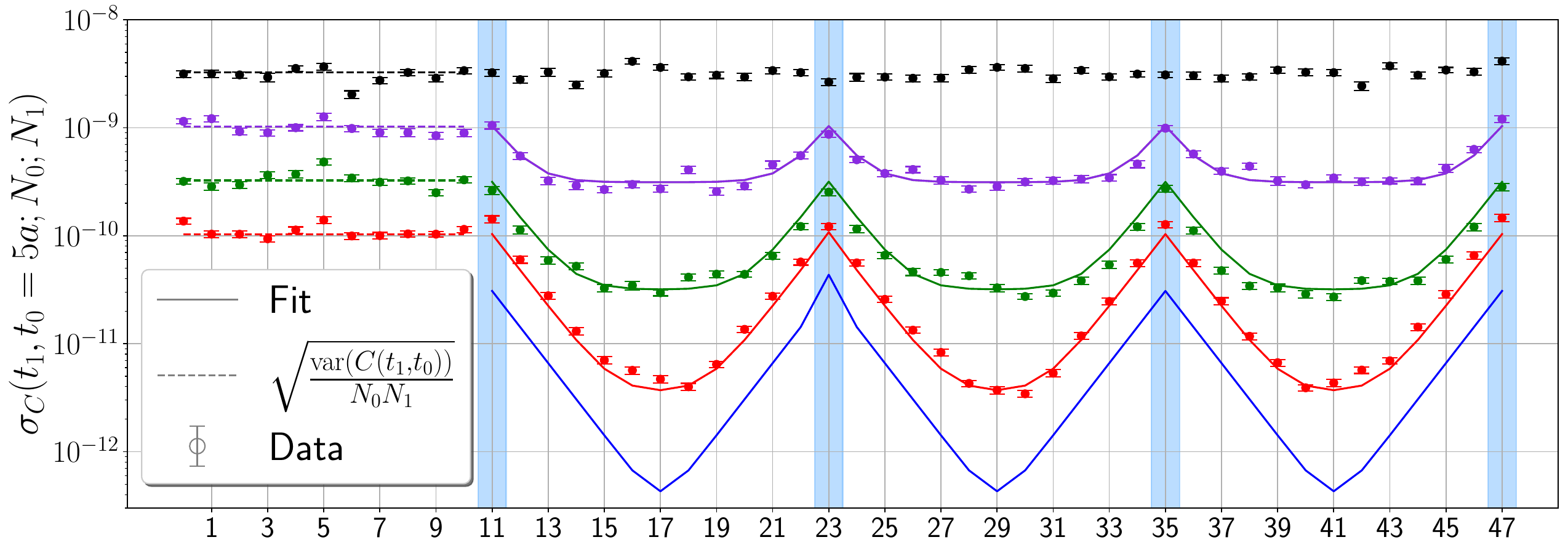}
	\end{subfigure}
	\begin{subfigure}[b]{0.85\textwidth}
		\centering
		\includegraphics[width=\textwidth]{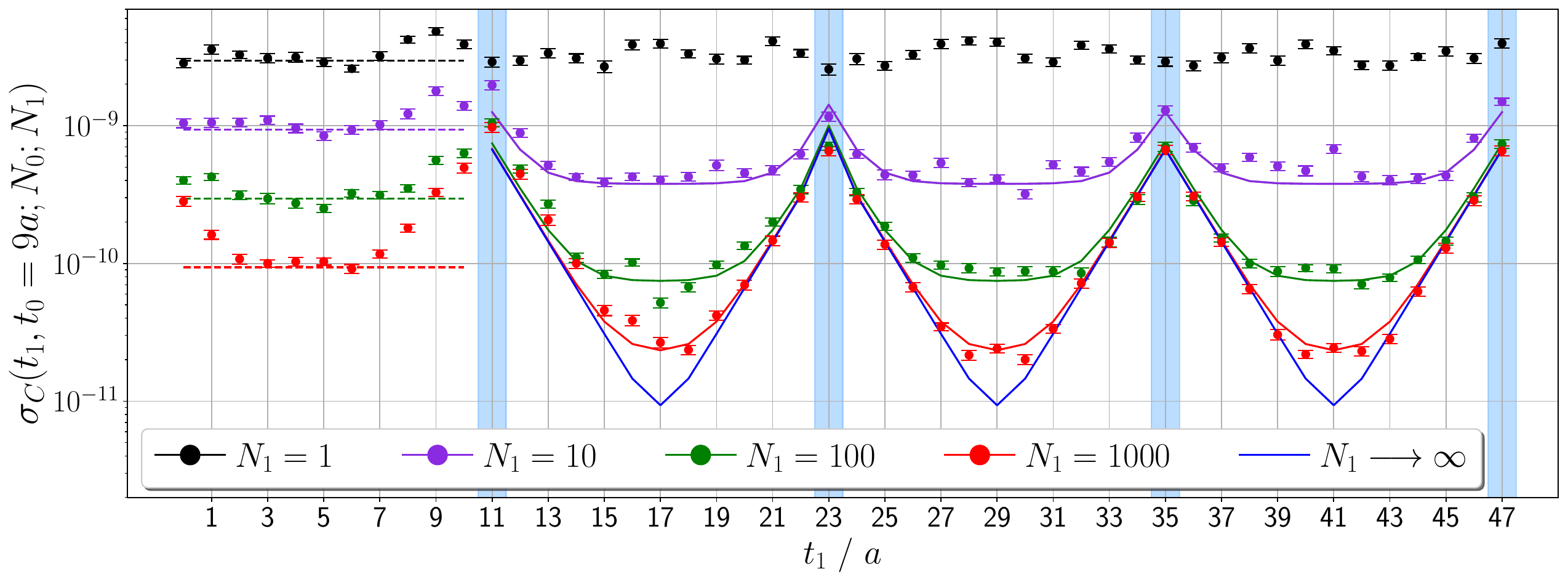}
	\end{subfigure}
	\caption{Error of a single $E^{++}$ glueball correlation function in the level-0 average at $t_0=5a$ (top), $9a$ (bottom), 
	and at different values of $t_1$, with the ensemble \textit{gb62} in Tab.~\ref{tbl:ensembles}.
	The solid lines represent the fit formula in Eq.~\eqref{mlvl_error} with $c_0 = c_1 = c_2$,
	where we consider all the nearest boundary corrections.
	The dashed lines are drawn only when the operators are in the same regions and correspond to the standard scaling in Eq.~\eqref{error_standard}.
	The light-blue shaded bands highlight the locations of the boundaries, see Fig.~\ref{Figure:dt1plus11}.
	The mass in the fit formula is set to twice the GEVP mass in the $E^{++}$ channel ($\widetilde{m}=2m^{E^{++}} \approx 1.54/a$), see Fig.~\ref{Figure:gb_meff_b62_b608}. 
	We add the $N_1 \to \infty$ lines for comparison, which correspond to the square root of the third term in Eq.~\eqref{mlvl_error}.}
	\label{fig:err-boundary-analytic}
\end{figure*}

Fig.~\ref{fig:err-boundary-analytic} shows the standard error of the correlation function $C^{\Gamma}(t_1, t_0)$ for $\Gamma=E^{++}$ at $t_0=5a$ (top), and  $t_0=9a$ (bottom) for different values of $t_1$. 
The source operator is fixed at $t_0$ while the sink operator at $t_1$ is moved 
to determine the dependence of the error on the distance of the sink from the boundaries.
The correlation function is computed on the ensemble \textit{gb62} using the level-1 decomposition $\Lambda_{11}$, see Fig.~\ref{Figure:dt1plus11}.
When the two operators are measured in different regions, the statistical uncertainty
follows a $cosh$ behavior as the operator moves between adjacent boundaries.
As the large $N_1$ limit is approached, the error decreases exponentially
according to $\exp(-m^\Gamma \Delta_1)\exp(-m^\Gamma \Delta_0)$ when $11\le t_1/a\le 23$.
In contrast, when the operators are in the same dynamical region
$(0\le t_1/a \le 10)$, statistical fluctuations are independent of separation, so the uncertainty falls only in inverse proportion to $\sqrt{N_1}$. Using
a global fit of the data to Eq.~\eqref{mlvl_error}, we find 
$c_0 \approx c_1, c_2$. This model behavior is compared to our data 
on the figure along with the limiting behavior when $N_1 \to \infty$. In this
limit, the third term in the expression, independent of $N_1$, persists, which 
suggests sub-sampling more than $N_1 \propto e^{m^\Gamma\Delta t}$ measurements does not improve precision any further. 

\begin{figure}[hpt!]
	\centering
	\includegraphics[width=0.48\textwidth]{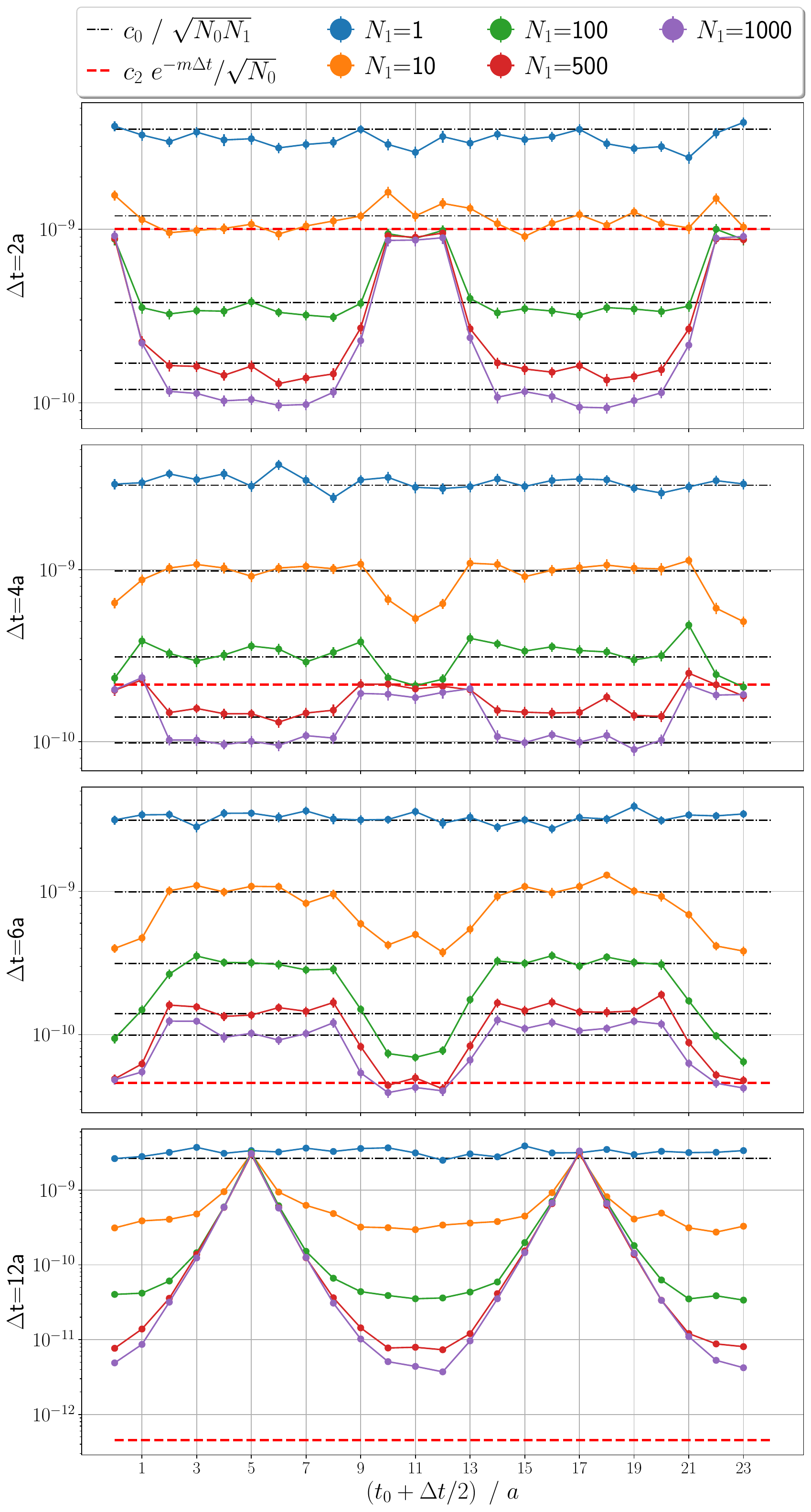}
	\caption{Errors of one $E^{++}$ glueball correlation function on the ensemble \textit{gb62} 
		at fixed values of $\Delta t=2a$, $4a$, $6a$, $12a$, with $\Delta t=t_1 - t_0$, 
		and with different values of $N_1$, $t_0$, and $t_1$. The thin black dashed lines represent the standard scaling $c_0/\sqrt{N_1}$ in Eq.~\eqref{error_standard}. The red dashed line stems from the third term in Eq.~\eqref{mlvl_error} where $\widetilde{m}=2m^{E^{++}}$. 
		This term is responsible for the saturation of the multi-level scaling at short distance. 
		Notice that at $\Delta t=2a$ and $\Delta t=4a$, the errors are larger when the two operators are in two different regions (e.g. $t_0+\Delta t/2=11a$) than when they are in the 
		same region (e.g. $t_0+\Delta t/2=3a$), even with $N_1=1000$. At $\Delta t=6a$, the multi-level scaling outperforms the standard scaling. 
		We display for clarity only half of the temporal extent of the lattice.
		To understand better where the operators are located, whether in the same region or not, one has to refer to Fig.~\ref{Figure:dt1plus11}.
	}
	\label{fig:err_fixed_t}
\end{figure}
The four panels of Fig.~\ref{fig:err_fixed_t} show the dependence of the error 
on the operator location relative to the boundaries for different values of 
$N_1$, for $\Delta t/a=2,4,6$ and $12$. As before, for certain values 
of 
$t_0$, the two operators are in the same sampling region, so the error is 
proportional to $1/\sqrt{N_1}$. In contrast, for values of $t_0$ where the operators
are in different regions, the error behaves according to Eq.~\eqref{mlvl_error}.
The x-axis is chosen so operators symmetrically distant from the boundaries 
are horizontally aligned, emphasizing the error reduction at different 
values of $t_0$. At short distance, the multi-level error saturates at $N_1
 \approx c_2\exp (m^\Gamma\Delta t)$. For the $E^{++}$ channel on ensemble 
\textit{gb62}, this corresponds to $\tilde{N}_1\approx22$ and 
$\tilde{N}_1\approx102$ for $\Delta t=4a,~6a$, respectively. Note the largest error reduction occurs when the two operators are at equal distances from the 
boundaries ($\Delta_1 = \Delta_0=\Delta t / 2$). This is seen in Fig.~\ref{fig:err_fixed_t}, where the multi-level error saturates at the red 
dashed line for $N_1 \gg \tilde{N}_1$. As a result, it is more efficient to
determine the correlation function at short distances from operator pairs in
the same dynamical region rather than adopting the multi-level since the 
standard error reduction, $\sqrt{N_1}$ is larger than the multi-level reduction
$\exp(m^\Gamma\Delta t)$ after saturation. Beyond $\Delta t/a=6$,
 Fig.~\ref{fig:err_fixed_t} shows that the multi-level scheme is more efficient 
than standard Monte Carlo. 
\subsection{Universal scaling toward the continuum limit}\label{Section:sec3_subsec1}
\begin{figure}[pt!]
	\centering
	\includegraphics[width=0.50\textwidth]{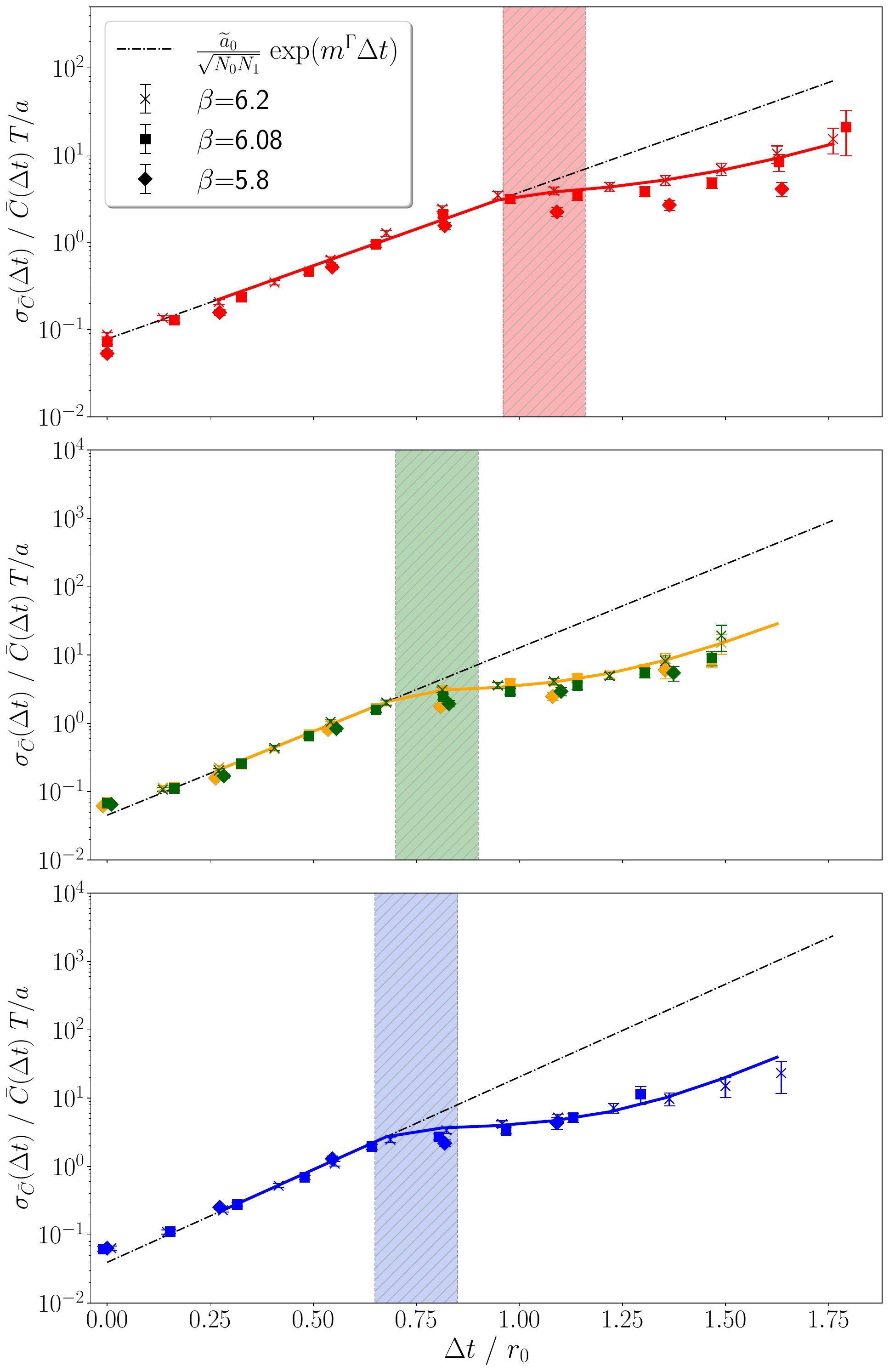}
	\caption{The noise-to-signal ratio exhibits a universal change of scaling in units of $r_0$ for the channels $A_1^{++}$ (red), 
		$E^{++}$ (green), $T_2^{++}$ (orange) and $A_1^{-+}$ (blue). 
		The dashed lines represent the standard scaling from the Lepage-Parisi argument \cite{Parisi:1983ae, Lepage:1989hd}.
		The solid lines are fits to the data at $\beta=6.2$ using Eq.~\eqref{ns_shortdistance} at short distance (left side of bands), and Eq.~\eqref{ns_longdistance} 
                (right side of bands) at long distance. The bands highlight approximately where the transition between the two error scalings occurs.}
	\label{Figure:universal_error_scaling_continuum}
\end{figure}

We investigate the scaling of the multi-level algorithm toward the continuum limit 
($\beta= 5.8, 6.08,$ and $6.2$) for the glueball channels $\Gamma = A_1^{++}$, $E^{++}$, $T_2^{++}$, and $A_1^{-+}$. For each correlator computed at different $t_1$ and $t_0$, we consider the weighted average in Eq.~\eqref{weighted_avrg}.
Fig.~\ref{Figure:universal_error_scaling_continuum} shows the noise-to-signal ratio of the weighted average glueball correlation 
functions in units of $r_0$ for a single glueball operator in all four channels. 
The data shows that at small distance ($\Delta t / r_0 \lesssim 0.6$) the noise-to-signal ratio increases exponentially like
$\sim \mathrm{exp}(m^{\mathrm{\Gamma}}\Delta t)$, which is the standard scaling
with traditional algorithms.
The reason is at short distance, the multi-level error reduction is not large, as observed in Fig.~\ref{fig:err_fixed_t}, 
so operators located in the same region contribute more to the weighted average, 
because their statistical errors are smaller.
It is important to mention that if one considers only operators located in two different dynamic regions, the resulting relative errors of the weighted average would be constant from $\Delta t=0a$ to $\Delta t=8a$ or $9a$ depending on the channel. This is because the multi-level error saturates at large $N_1$ at a value of $\sigma_C \propto \mathrm{exp}(-m^\Gamma \Delta t)$,
which is also how the signal drops over time.
At intermediate time separations ($0.6r_0 \le \Delta t \le 1.1r_0$), there is a change of scaling where the noise-to-signal ratio is almost constant.
This region is highlighted by the shaded blue band in Fig.~\ref{Figure:universal_error_scaling_continuum}.
The physical distance where the transition occurs is universal for each channel, as expected from Eq.~\eqref{transition_point} 
because $\mathrm{exp}(m^\Gamma\Delta t)$ does not depend on $\beta$.
The dependence on the mass can be observed between the $A_1^{++}$ channel and the other channels, 
the former contains the lightest state in the spectrum, thus requiring slightly larger time separations for the change of scaling.
After this transition, the noise-to-signal ratio follows the multi-level scaling, i.e. the square root of Eq.~\eqref{ns_longdistance},
which is observed until the error of the error becomes too large.
Thus, to keep the relative error constant over time at large separations, one has to set $\tilde{N}_1 \propto \mathrm{exp}(m^\Gamma\Delta t)$.
%
\subsection{Spectrum results}\label{Section:sec3_subsec2}
To determine the glueball spectrum, we construct a basis consisting of Wilson loops with $35$ different shapes,
as discussed in Appendix~\ref{appendix:app1}, some of which can be projected onto the channels 
of interest $\Gamma = A_1^{++}$, $E^{++}$, $T_2^{++}$, and $A_1^{-+}$ as shown in \citep{Berg:1982kp, Georgi2018}. 
Each of these Wilson loops is built from links smeared with $5$ different APE smearing levels \cite{Falcioni:1984ei} 
to construct an efficient variational basis made of up to $35\times5$ operators.

\begin{figure}[ht]
	\centering
	\includegraphics[width=0.5\textwidth]{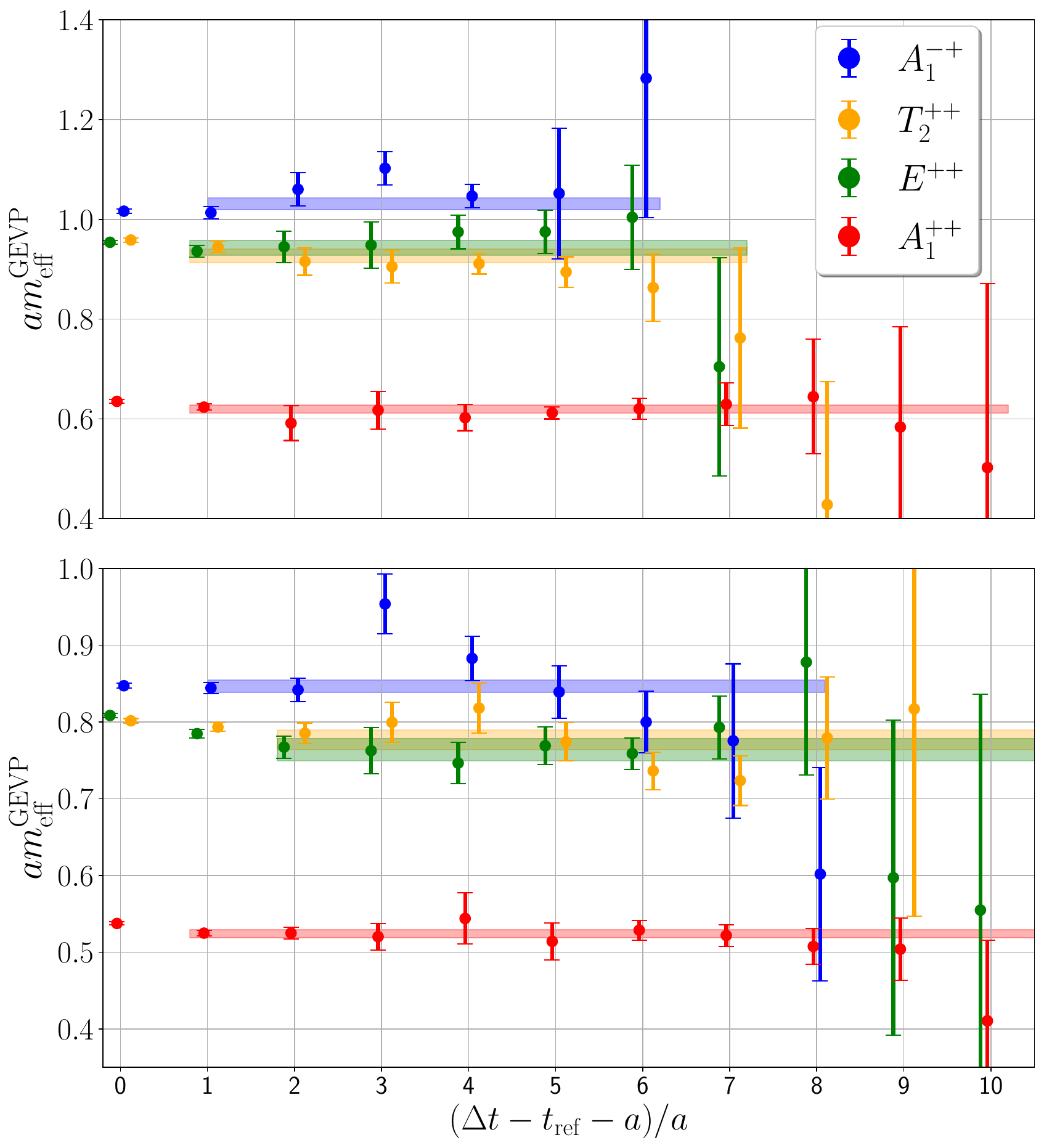}
	\hfill
	\caption{(top) GEVP glueball masses at $\beta=6.08$ for all four channels in the legend;
		(bottom) GEVP glueball masses at $\beta=6.2$ for all four channels.}
	\label{Figure:gb_meff_b62_b608}
\end{figure}

\begin{figure*}[ht]
	\centering
	\includegraphics[width=0.93\textwidth]{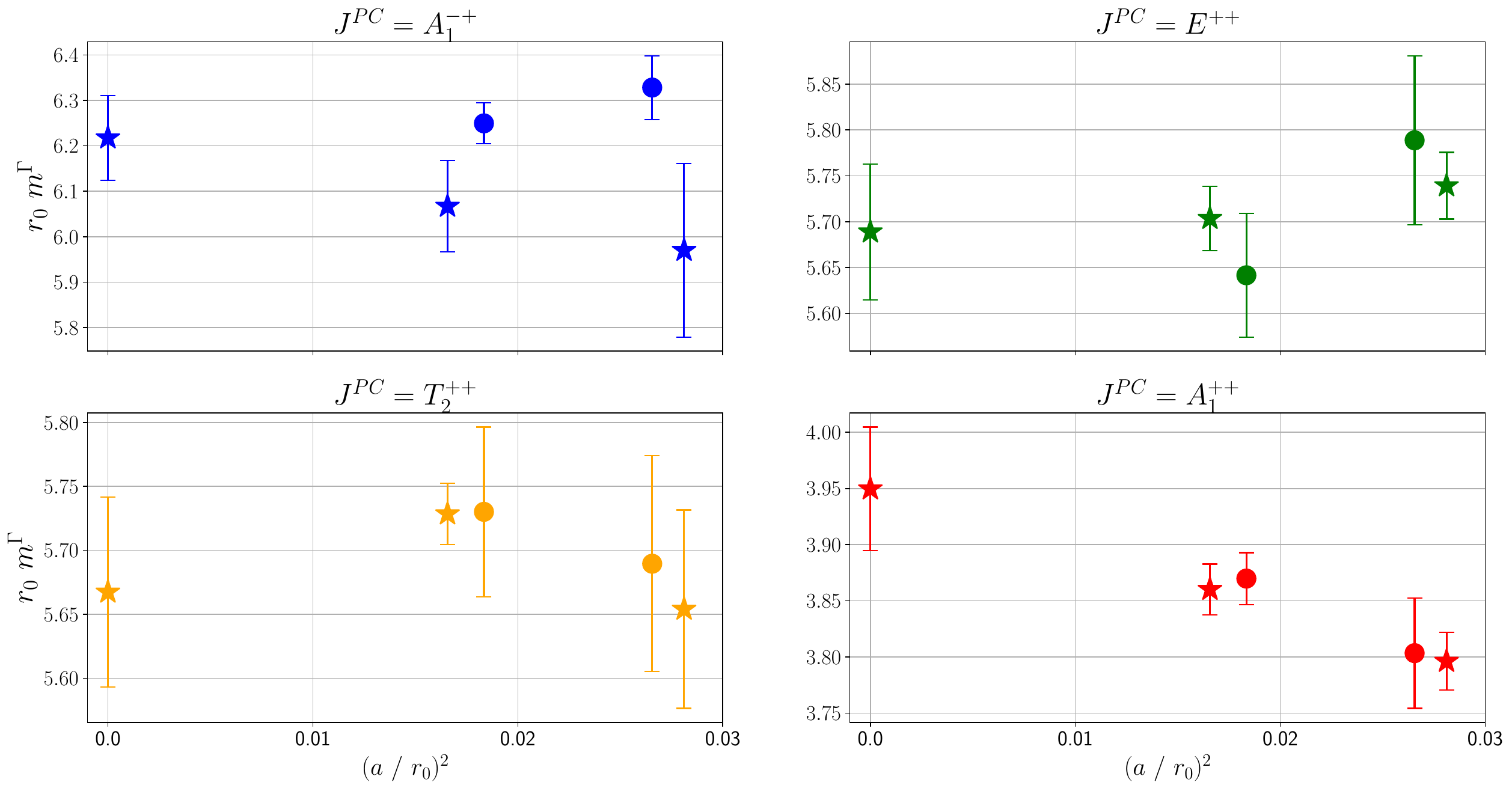}
	\caption{Comparison between glueball masses in this work at $\beta=6.2, ~ 6.08$ with multi-level algorithm (circles), and state-of-the-art results (stars) at $\beta=6.0625, ~6.235$ 
		and continuum limit results of \cite{Athenodorou:2020ani}, where the traditional Monte Carlo algorithm is adopted.}
	\label{Figure:gb_meff_r0_continuum}
\end{figure*}

For each operator $\oo^{\Gamma}(t)$ in this basis, we estimate the matrix of correlation functions
\begin{equation}
	C^{\Gamma}_{\alpha \beta}(t_1, t_0) = 
	\langle \left[\oo^{\Gamma}_\alpha( t_1)\right] \left[\oo^{\Gamma}_\beta(t_0)\right] \rangle ~,
	\hspace{1.2em}
	\alpha,\beta = 1,\dots, N_{\Gamma}~,
	\label{eqn:CorrelationMatrix}
\end{equation}
with a total of $N_1 = 1000$ sub-measurements and $N_{\Gamma}$ denotes the number of distinct operators formed from combinations of Wilson loops projected 
onto irrep $\Gamma$. 
We estimate the weighted average $\overline{C}^{\Gamma}_{\alpha \beta}(\Delta t)$ 
as discussed in the previous section, see Eq.~\eqref{weighted_avrg}. 
We then solve for $\Delta t > t_{\mathrm{ref}}$ the GEVP
\begin{align}\label{gevp}
\begin{split}
	\overline{C}^{\Gamma}(\Delta t) V^{\Gamma}(\Delta t, t_{\mathrm{ref}}) &= 
	\overline{C}^{\Gamma}(t_{\mathrm{ref}})  V^{\Gamma}(\Delta t, t_{\mathrm{ref}}) \Lambda^{\Gamma}(\Delta t, t_{\mathrm{ref}})
	~,
\end{split}
\end{align}
where $\Lambda^{\Gamma}=\mathrm{diag}\left(\lambda^{\Gamma}_1(\Delta t, t_{\mathrm{ref}}), \dots, \lambda^{\Gamma}_{N_{\Gamma}}(\Delta t, t_{\mathrm{ref}}\right)$ 
and $V^{\Gamma}=\left(v^{\Gamma}_1(\Delta t, t_{\mathrm{ref}}), \dots, v^{\Gamma}_{N_{\Gamma}}(\Delta t, t_{\mathrm{ref}})\right)$ are the matrices of generalized eigenvalues and eigenvectors, respectively.
The GEVP effective masses of ground state and further excitations $m^{\Gamma}_{k}$ can be extracted from the eigenvalues 
as $\lambda^{\Gamma}_k(\Delta t, t_{\mathrm{ref}}) \propto  \mathrm{exp}\left({-m^{\Gamma}_{k}{(\Delta t- t_{\mathrm{ref}})}}\right)$
for the channels 
$\Gamma=A_1^{++}, E^{++}, T_2^{++}, A_1^{-+}$ \citep{Luscher, Blossier}. 
Some operators with different shapes might be degenerate 
at large enough smearing radii, as observed in Fig.~8 of \cite{Sakai:2022zdc}, and thus some careful choice must be made for the optimal variational basis.
Thus, we do not employ the full operator basis for each channel and, in practice, use $\mathcal{O}(10)$ operators for each glueball channel. 
We adopt the two-level algorithm on $N_0=101$ configurations with up to $N_1=1000$ sub-measurements to compute these correlation functions and solve the GEVP. 
In Fig.~\ref{Figure:gb_meff_b62_b608}, we present results at $\beta=6.2, ~6.08$ for the ground state of the different channels.
The small irregularity at $\Delta t-t_{\mathrm{ref}}-a = 3a$ in the $A_1^{-+}$ channel is a statistical fluctuation 
and is due to the loss of translational invariance of the error after combining multi-level with standard measurements. 
In particular, if the statistical error is slightly underestimated, 
it will enhance some correlators in the weighted average, as the weights are the inverse of the variance, see Eq.~\eqref{weighted_avrg}.
A comparison with state-of-the-art results of \cite{Athenodorou:2020ani} at similar $\beta$ values but slightly different volumes shows that the multi-level results have smaller errors at large source-sink separations due to better scaling, as shown in Fig.~\ref{Figure:universal_error_scaling_continuum}.
In particular, in Fig.~\ref{Figure:gb_meff_r0_continuum}, we show a comparison between our multi-level results at $\beta=6.2$, $6.08$ 
and the state-of-the-art results in \cite{Athenodorou:2020ani} at $\beta=6.235$, $6.0625$. 
The results agree within the errors, and the statistical uncertainty is also 
very similar.
The statistics used in the two calculations are roughly the same ($N_0 \times N_1 = \mathcal{O}(10^5)$), along with the computational cost of all the updates ($\mathcal{O}(10^6)$ MDUs),
and we also adopt a large variational basis.
Although the multi-level improves the noise-to-signal at large distance, c.f. Fig.~\ref{Figure:universal_error_scaling_continuum},
the algorithm does not improve the final uncertainty of the glueball effective masses.
Glueball masses are extracted from a fit that starts at short distance, 
where the multi-level does not work efficiently. This explains why the uncertainty for the final value 
of the glueball effective masses is roughly similar to \cite{Athenodorou:2020ani}, see Fig.~\ref{Figure:gb_meff_r0_continuum}.
However, the multi-level signal is determined to much longer distances where the ground state dominates, which gives more confidence in the reliability of the plateau estimated at short distance.
The multi-level algorithm will be necessary for investigations of glueball mixing with quark-anti-quark mesons when dynamical fermions are included as the signal drops quickly
and it is challenging to understand the gluonic content of the mixed 
states \cite{Urrea-Nino:2023ysv}.

%% file: sections/sec4.tex
\section{Conclusions}\label{Section:sec4}

We have investigated the efficiency of a two-level algorithm for determining the glueball mass spectrum in various channels toward the continuum limit. It was not a priori clear that the method would be efficient.
As we demonstrated, the standard estimator outperforms the two-level measurements at short distances. This is primarily due to the breaking of the translational invariance by the fixed boundaries and the effect of the fixed fields on them, which limit the exploration
of field space in their vicinity.
With the standard method, the noise-to-signal ratio deteriorates exponentially, and the signal could be lost before the 
two-level estimator becomes efficient. This is complicated further because state-of-the-art analyses combine correlation functions at various source-sink separations using a sizeable variational basis with a GEVP to extract effective masses. We have demonstrated that in this setting the two-level method works. The deterioration is stopped, and the noise-to-signal ratio stays constant for a certain time interval.

The results agree very well with state-of-the-art analyses that
adopt the traditional Monte Carlo. The overall statistical uncertainty is not
improved; however, the GEVP masses are extracted from fits starting at short
time separation, where the two-level is ineffective. Still, with the same
cost as traditional Monte Carlo, the two-level algorithm can render
correlation functions at long distance with an exponentially reduced error, giving more confidence in the plateau estimate.

Depending on the channel, the two-level method is more efficient for
source-sink separations above $0.3$~fm to $0.6$~fm, and we can extend the
plateau region by roughly $0.25$~fm using $N_1=1000$ sub-measurements. We can
reliably model the performance of the algorithm depending on the source and
sink position as well as the resulting weighted averages. This allows us to understand
the transition in the dominance in the average from the standard to the two-level measurements with increasing
time.

The two-level Monte Carlo is an expensive method. To offset the reduced
averaging in time due to the fixed regions, the number of sub-measurements needs
to be large, where the full power is only reached at significant source-sink
separations. To ensure a reliable error analysis, boundary fields must be sampled with at least $\mathcal{O}(10^2)$ measurements. 
This requirement translates to a cost equivalent to taking measurements on $\mathcal{O}(10^5)$ gauge field 
configurations for $N_1=1000$.

Our study has been performed in pure gauge theory, where an operator basis can be found that allows
plateau fits from very short time separations. In the presence of fermions, this is no longer possible. The large separations that can be reached with the two-level algorithm will become essential, for instance, in studies of glueball-charmonium mixing. Here standard algorithms cannot reach trustworthy plateau regions such that reliable computations need improved methods like the one investigated here.

\acknowledgements
The work is supported by the German Research Foundation (DFG) research unit FOR5269
"Future methods for studying confined gluons in QCD." The authors thank all the members
of the research unit for useful discussions. L. B. is grateful to J. Frison and A. Risch for discussions. 
S.~M. received funding from the European Union’s Horizon 2020 research and
innovation program under the Marie Sk\l odowska-Curie grant agreement \textnumero 813942.
This work was supported by the
STRONG-2020 project, funded by the European Community Horizon 2020 research and 
innovation programme under grant agreement 824093. 
The authors gratefully acknowledge the scientific support and HPC resources provided by the
Erlangen National High Performance Computing Center (NHR@FAU) of the
Friedrich-Alexander-Universit\"at Erlangen-N\"urnberg (FAU) under the NHR project k103bf.

%% file: appendices/appendix1.tex
\section{Construction of the glueball operators}\label{appendix:app1}
\begin{figure}[htbp!]
	\centering
	\begin{tabular}{ccc}
		\includegraphics[width=0.05\textwidth]{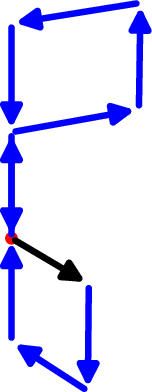} &   \includegraphics[width=0.12\textwidth]{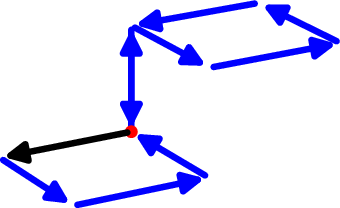} &
		\includegraphics[width=0.09\textwidth]{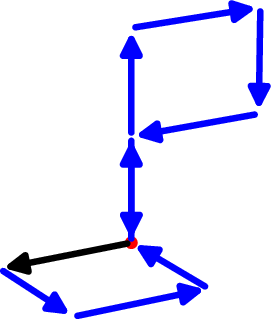} \\ \includegraphics[width=0.09\textwidth]{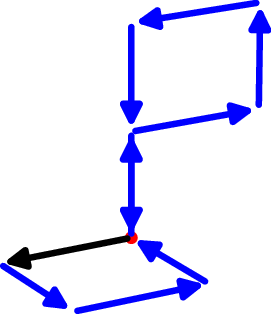} &
		\includegraphics[width=0.09\textwidth]{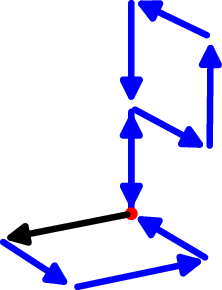} &
		\includegraphics[width=0.09\textwidth]{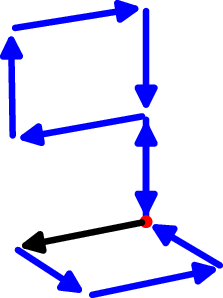} \\ \includegraphics[width=0.09\textwidth]{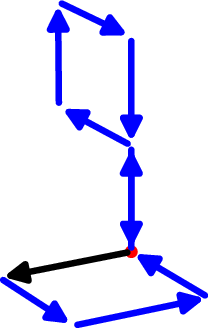} & \includegraphics[width=0.12\textwidth]{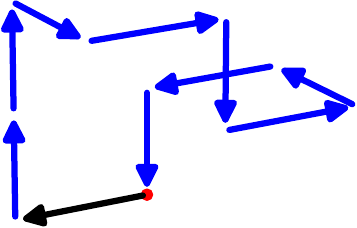} & \includegraphics[width=0.12\textwidth]{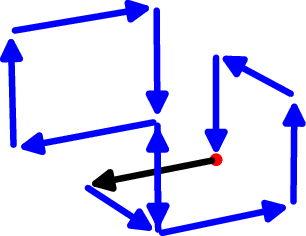} \\ \includegraphics[width=0.15\textwidth]{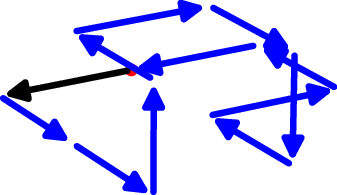} &
		\includegraphics[width=0.15\textwidth]{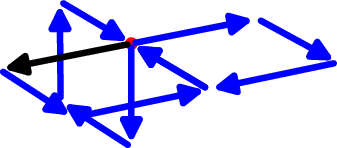} & \includegraphics[width=0.15\textwidth]{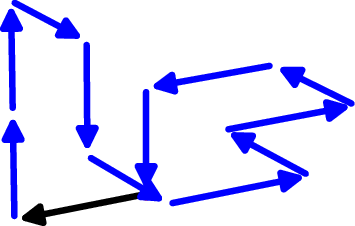} \\ \includegraphics[width=0.09\textwidth]{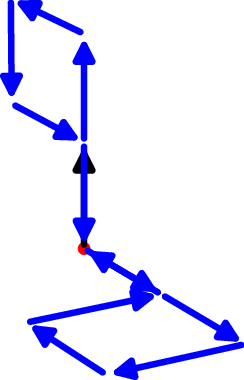}
	\end{tabular}
	\caption{Additional loop shapes used to increase the number of glueball operators.}
	\label{fig:NewShapes}
\end{figure}

The starting point to build the glueball operators used in this work is a given loop shape, e.g. the 22 
different ones presented in \cite{Berg:1982kp} of lengths 4, 6 and 8. A 3D Wilson loop of fixed shape starting 
at spatial point $\vec{x}$ and Euclidean time $t$ is denoted as $W^{s}\left( \vec{x}, t  \right)$, where the 
index $s$ labels the shape. Elements of the cubic group can 
act on such a Wilson loop when it is represented as a tuple of unitary displacements 
equivalent under cyclic permutations thanks to the projection to zero spatial momentum \citep{Berg:1982kp}. The loop obtained after acting with group element $g_l$, $l=0,...,23$, on $W^{s}\left( \vec{x}, t  \right)$ is denoted as $W^{s}_l\left( \vec{x}, t  \right)$. Considering only the $l$ index and not any degeneracies between loops, these 24 loops form a basis which generates the regular (or permutation) representation of the cubic group, and therefore there exist linear combinations of them which transform according to any of the five irreps of this group \cite{Georgi2018, Bunker1998-dj}. 
A glueball operator with zero spatial momentum which transforms according 
to a given irrep $R$ can be written as
\begin{align}
W^{\left( R, k, s \right)}(t) &= \frac{a^3}{L^3} \sum_{\vec{x}} \sum_{l=0}^{23} c_l^{R, k} \text{Tr}\left[ W^{s}_l\left( \vec{x}, t  \right)\right]~,
\end{align}
where $k=0,...,\text{dim}(R)-1$ denotes the copy of irrep $R$ which appears in the regular representation and $c_l^{R, k}$ 
are the projection coefficients coefficients for copy $k$ of irrep $R$. These coefficients can be calculated 
via projection methods \cite{Georgi2018, Bunker1998-dj}. In this work we consider only one copy per irrep so we omit the $k$ index from now on. Parity is fixed by considering the sum $(P=+1)$ or difference $(P=-1)$ of each loop with its parity twin, i.e. the loop under the action of a parity transformation. The parity twin of $W^{s}_l\left( \vec{x}, t  \right)$ will be denoted as $W^{s}_l\left( \vec{x}, t  \right)^{P}$. Charge conjugation symmetry is fixed by taking the real $(C = +1)$ or imaginary $(C=-1)$ part of the trace. A glueball operator $\mathrm{O}^{ \Gamma}_s(t)$ which transforms according to a fixed irrep $\Gamma = R^{PC}$, 
is given by
\begin{align}
\begin{split}
	\mathrm{O}&^{ R^{\pm +} }_s(t) \\&= \frac{a^3}{L^3} \sum_{\vec{x}} \sum_{l=0}^{23} c_l^{R} 
	\text{Re}\left\{ \text{Tr}\left[ W^{s}_l\left( \vec{x}, t  \right) \pm W^{s}_l\left( \vec{x}, t  \right)^{P}  \right]\right\}~,
\end{split}
\\
\begin{split}
	\mathrm{O}&^{ R^{\pm -} }_s(t) \\&= \frac{a^3}{L^3} \sum_{\vec{x}} \sum_{l=0}^{23} c_l^{R} 
	\text{Im}\left \{ \text{Tr}\left[ W^{s}_l\left( \vec{x}, t  \right) \pm W^{s}_l\left( \vec{x}, t  \right)^{P}  \right]\right \} ~.
\end{split}
\end{align}
Since this construction is independent of the link smearing use, we can introduce a single index $\alpha$ which accounts for both the loop shape and smearing used to define 
a given glueball operator in the basis. In this manner, the glueball operators are denoted as $\oo^{\Gamma}_\alpha(t)$ 
as used in Eq.~\eqref{eqn:CorrelationMatrix}. Once the projection coefficients are known, operators for all 20 
symmetry channels can be built using the above expressions for a choice of shape; however, several of them could yield zeros. 
This is because not all irreps of the cubic group extended to include parity and charge conjugation are contained in the 
reducible representation which a given shape generates. It is the effect of taking into account the possible degeneracies between the loops under the action of the extended cubic group, which can reduce the dimension of the generated representation. For example, the plaquette cannot be used to build an operator 
which transforms according to any $T_2^{PC}$. To avoid getting these zeros, one first checks which irreps are accessible 
for each loop shape and only builds their corresponding operators. Tables including this irreducible content for loops of up to length 8 are given in \cite{Berg:1982kp}. As a final remark, only 5 out of the 22 shapes presented 
in \cite{Berg:1982kp} can be used to access the $A_1^{-+}$ irrep. To increase the number of operators for this channel, 
we use 13 additional loop shapes shown in Fig.~\ref{fig:NewShapes}. Out of these, 8 have length 10 and 5 have length 12. 
They are chosen such that they all can be used to access the $A_1^{-+}$ irrep as well as the $A_1^{++}$, $E^{++}$ and $T_2^{++}$.
